\documentclass[%
 reprint,
 amsmath,amssymb,
 aps,
pra,
floatfix,
]{revtex4-2}

\usepackage{graphicx}
\usepackage{dcolumn}
\usepackage{bm}
\usepackage{color}

\usepackage{subfigure}
\usepackage{amsmath}
\usepackage{physics}
\usepackage{caption}
\usepackage{subcaption}
\usepackage[maxfloats=256]{morefloats}
\usepackage{hyperref}
\usepackage{multirow}

\begin{document}

\preprint{APS/123-QED}

\title{Heisenberg spin networks for realizing quantum battery with the aid of Dzyaloshinskii–Moriya interaction}
\author{Suprabha Bhattacharya$^{1}$}
\email{m23ph1018@iitj.ac.in}
\author{Vivek Balasaheb Sabale$^{1}$}
\email{sabale.1@iitj.ac.in}
\author{Atul Kumar$^{1}$}
\email{corresponding author: atulk@iitj.ac.in}
\affiliation{$^{1}$ Indian Institute of Technology Jodhpur, 342030, India}

\begin{abstract}
This work investigates the energy storage properties of quantum spin chains in the context of quantum batteries by introducing Heisenberg spin \textcolor{black}{network} models organized into different configurations: open, closed, supercube geometries, and $c$-regular graphs. The charging dynamics of these systems are examined using Hamiltonians that include contributions from the battery, spin–spin interactions, and a transverse magnetic field. Incorporating the Dzyaloshinskii–Moriya interaction (DMI) into the charging Hamiltonian is found to enhance the ergotropy in the XXZ model, particularly for the supercube configuration, thereby improving quantum battery performance. To explore the role of structural variations, we extend our study to $c$-regular graphs with system sizes ranging from $3$ to $12$ qubits, including highly symmetric geometries such as the tetrahedron, octahedron, and icosahedron. These analyzes reveal that such symmetric structures retain ideal sinusoidal charging–discharging behavior when DMI is tuned appropriately, establishing symmetry and coordination as key principles for scalable quantum battery architectures.
\end{abstract}

\keywords{Quantum Spin Chain, Quantum Battery, Dzyaloshinskii-Moriya interaction} 
\maketitle


\section{Introduction}

\begin{figure*}[t]
    \centering
    \includegraphics[width=0.75\linewidth]{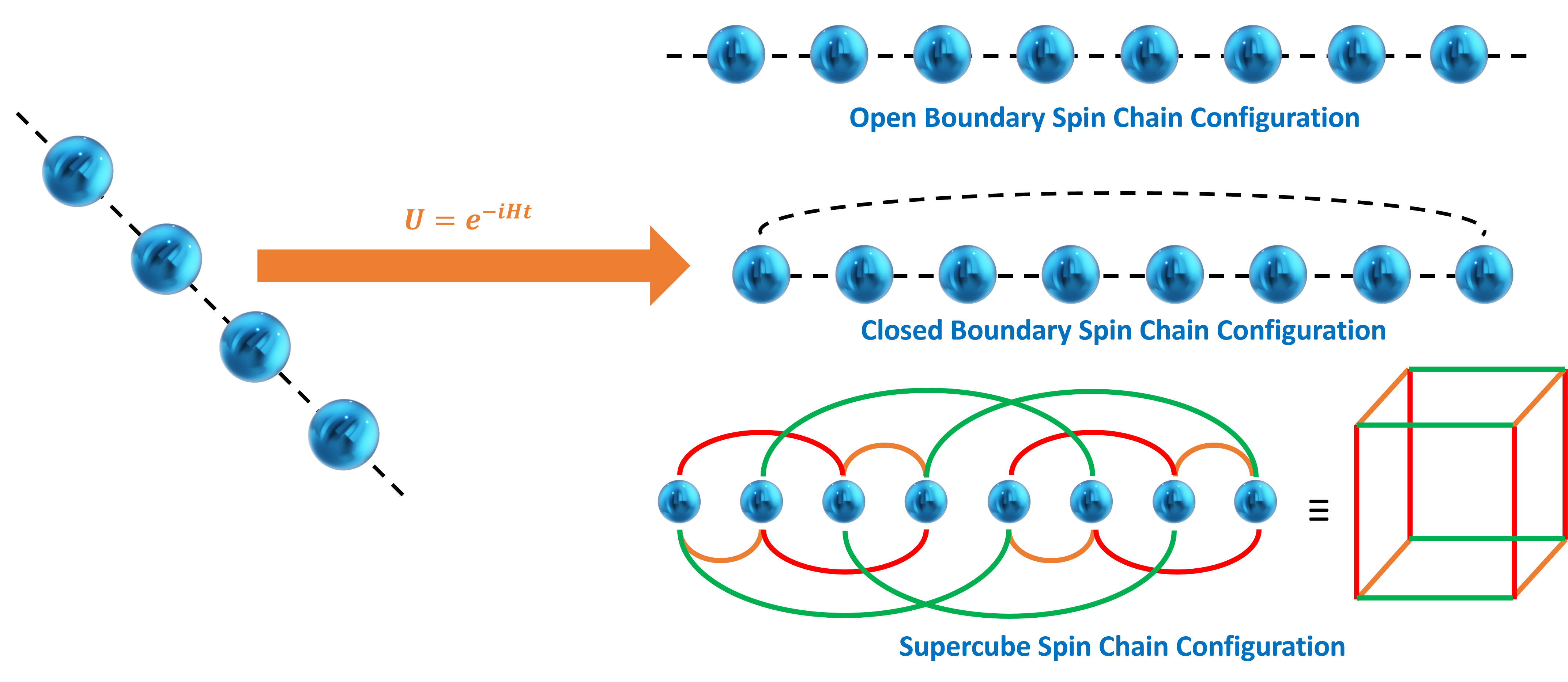}
    \caption{Visualizing possible inter-qubit interaction and connections developed during the evolution. }
    \label{fig:config}
\end{figure*}
As classical electronics approach atomic-scale limits, quantum effects have become unavoidable in modern semiconductor devices operating at $3$–$5$ nanometers. This fundamental shift challenges conventional design and opens new avenues for quantum-enhanced technologies. Future technologies must therefore account for and utilize quantum phenomena such as superposition, entanglement, and interference \cite{Horodecki2009} to exceed the capabilities of classical systems. Recent advances in the control of small quantum systems have facilitated the convergence of quantum information science \cite{nielsen2010quantum} and thermodynamic principles, leading to the emergence of quantum thermodynamics \cite{Vinjanampathy01102016, binder2018thermodynamics} as a distinct research domain. \par
A key application of quantum thermodynamics is the design of quantum batteries (QBs)- energy storage devices composed of quantum systems. The concept was first proposed by Alicki and Fannes \cite{Alicki2013}, and has since evolved through studies exploring the role of quantum coherence, many-body interactions, and collective charging in enhancing charging power and extractable work \cite{bhattacharjee2021quantum, quach2023quantum, binder2018thermodynamics, Campaioli_2017, Ferraro2018, Quantum_vs_classical, global_ops, energy_transfer, Farre2020, Modi2011, Giorgi_2015}. QBs are typically charged through unitary operations, enabling minimal entropy production and potentially offering a quantum advantage over classical batteries. Additionally, recent experimental progress in constructing a nanographene-based spin-$\frac{1}{2}$ chain with tunable length \cite{zhao2025spin} motivates the study of models that may be experimentally realizable. \par
Despite this progress, two critical gaps remain in the theoretical design of quantum batteries. Most existing studies focus on linear spin chains, central spin configurations, or all-to-all coupling schemes, with limited exploration of complex connectivity patterns \cite{Sachdev1993,Rossini2020, kim2022}. Second, the role of anisotropic interactions- particularly the Dzyaloshinskii–Moriya interaction (DMI), which is known to induce spin canting and enhance entanglement- has not been adequately studied in structured spin networks. While some works have incorporated DMI in basic QB models, its combined influence with interaction topology on charging dynamics in scalable architectures~\cite{many-body-QB} remains largely unaddressed. In this work, we address these gaps by constructing and analyzing Heisenberg spin \textcolor{black}{network} quantum battery models across a variety of configurations. We begin with eight qubits arranged in three representative topologies- open chain, closed chain~\cite{Zheng2025}, and a ''supercube'' structure- as illustrated in Fig.~\ref{fig:config}. We then extend our study to $c$-regular graphs with system sizes ranging from 3 to 12 qubits, including highly symmetric geometries such as the tetrahedron, octahedron, and icosahedron. In all cases, we incorporate anisotropic Heisenberg interactions and DMI into the charging Hamiltonian and evaluate performance using two established metrics: ergotropy (maximum extractable work) and charging power. Our approach builds on the spin chain model introduced in \cite{many-body-QB}, focusing specifically on antiferromagnetic Heisenberg spin chain Hamiltonians to construct QBs. Quantum batteries can be charged using either the parallel charging mode, where local fields act independently on each qubit, or the collective charging mode, where an interaction Hamiltonian acts globally on the system \cite{Binder2015, Kamin_2_3_qb}. Prior studies have shown that collective charging provides a quantum advantage over parallel charging \cite{Quantum_vs_classical, PhysRevResearch.4.043150, Peng2021, global_ops, kim2022}. \par
Our results show that the supercube configuration, under specific interaction strengths (Heisenberg coupling $J=1$, DMI strength $D=1.7$), achieves a near-ideal sinusoidal charging-discharging cycle with negligible residual energy. By contrast, open and closed chains display reduced performance and irregular charging patterns, particularly in the presence of DMI. We further analyze the role of system size by varying the number of qubits and by enriching connectivity through additional couplings (e.g., body and face diagonals). These studies reveal that interaction strengths must be carefully tuned to preserve periodic and efficient energy transfer. Notably, highly symmetric geometries such as Platonic solids yield results comparable to or surpassing the supercube, highlighting the importance of uniform connectivity in stabilizing charging dynamics.  To systematically probe scalability, we extend our study to a broad class of $c$-regular graphs with system sizes ranging from 3 to 12 qubits. This framework encompasses both generic regular connectivities and highly symmetric polyhedral skeletons, including the tetrahedron, octahedron, and icosahedron. At suitable DMI strengths, all of these structures exhibit ideal sinusoidal charging--discharging cycles, establishing symmetry and coordination as central design principles for scalable quantum batteries. \textcolor{black}{We additionally provide a more detailed analysis of how the DM interaction influences the population dynamics of the eigenstates and how this behavior depends on the underlying geometry, presented in the newly added Appendix section. This expanded analysis further allows us to confirm that the observed charging behavior does not originate from specific initial states or limited parameter choices; rather, it helps diagnose whether features such as enhanced charging arise from interaction-driven dynamics instead of entanglement effects. Placing these results in the Appendix enables readers to explore the full dynamical behavior without interrupting the flow of the main discussion, while still keeping all supporting analyzes readily accessible.} These findings indicate that charging efficiency, stability, and periodicity can be systematically engineered through interaction geometry, providing a new design axis for quantum energy storage systems and pointing toward strategies for scaling QBs beyond small sizes. \par


The structure of the article is as follows: \textcolor{black}{Section~\ref{sec:2} presents the proposed models}, including their Hamiltonians and geometries (Section~\ref{sec:3}), and performance metrics (Section~\ref{sec:3b}). \textcolor{black}{Section~\ref{sec:4} presents the results and comparative analysis across various models. Additionally, Appendix~\ref{A1} provides an analysis of population dynamics with and without DMI.} The study concludes in Section~\ref{sec:5}.

\section{Model and Performance Metrics for Quantum Battery \label{sec:2}}

 \begin{figure*}[ht]
    \centering
    \subfigure[Open boundary spin chain]{\includegraphics[width = 0.3\linewidth]{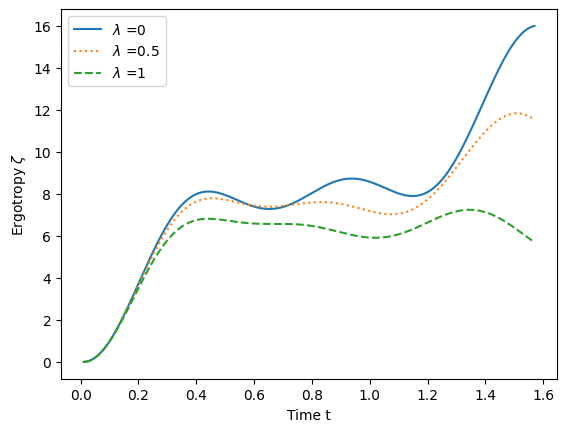}} 
    \subfigure[Closed boundary spin chain]{\includegraphics[width = 0.3\linewidth]{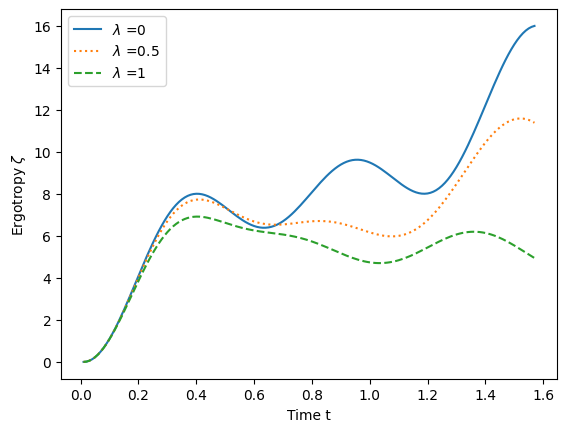}} 
    \subfigure[Supercube spin chain]{\includegraphics[width = 0.3\linewidth]{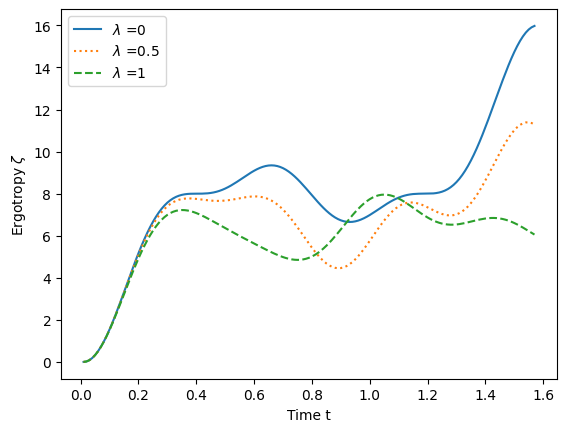}}
    \caption{Ergotropy for Ising model ($\delta = 1, \Delta = 0$) without DM interaction ($D = 0$)}
    \label{fig:erg ising}
\end{figure*}

\begin{figure*}[ht]
     \centering
     \subfigure[Open boundary spin chain]{\includegraphics[width = 0.3\linewidth]{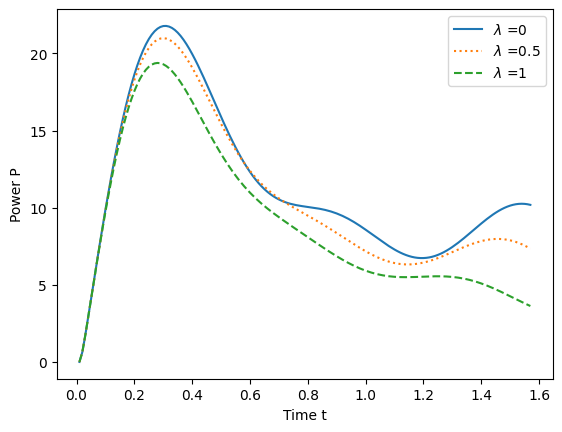}}
     \subfigure[Closed boundary spin chain]{\includegraphics[width = 0.3\linewidth]{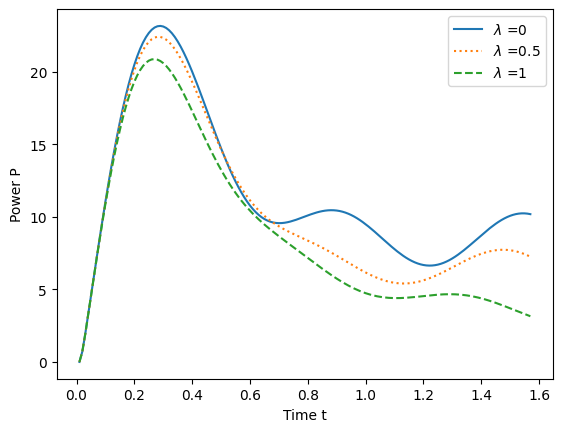}} 
     \subfigure[Supercube spin chain]{\includegraphics[width = 0.3\linewidth]{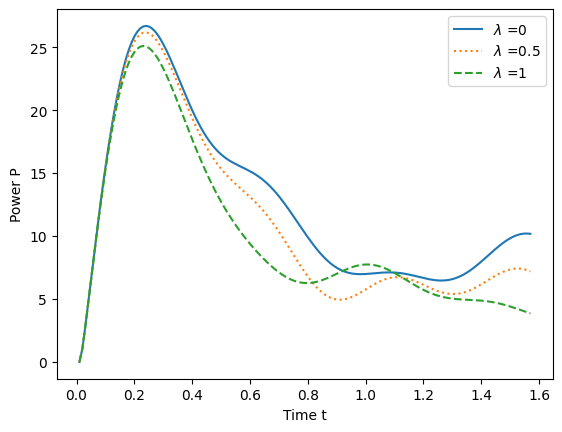}}
     \caption{Power for Ising model ($\delta = 1, \Delta = 0$) without DM interaction ($D = 0$)}
     \label{fig:power ising}
 \end{figure*}

\textcolor{black}{We now construct a model of a quantum battery based on interacting spin chains, which can be viewed as one-dimensional arrays of qubits. Such systems have been extensively studied in condensed matter physics \cite{auerbach2012interacting, blundell2001magnetism, kittel2018introduction}, quantum materials \cite{zhao2025spin}, and quantum communication and computation \cite{Zueco2009, Smith2019}. We follow the conventions outlined in Ref.~\cite{zhang2024quantum} to define the Hamiltonians relevant to our model.} Our objective is to investigate how collective effects arising from Heisenberg exchange and Dzyaloshinskii–Moriya interactions influence the charging and discharging performance of the battery. To this end, we consider an eight-qubit ($n=8$) spin chain as the working medium \cite{many_body_localized, many-body-QB, XYZ_chain, Dou2022, Huangfu2021, Peng2021, Arjmandi2022, Ghosh2022}.  \par

In the absence of charging, the system evolves under the internal battery Hamiltonian
\begin{equation}
    H_B = \hbar\omega_0 \sum_{i = 1}^{n} \sigma_i^z,
    \label{H battery}
\end{equation}
where $\omega_0$ is the uniform Larmor frequency, which determines the intrinsic energy level spacing of the battery, and $\sigma_i^z$ is the Pauli-$z$ operator acting on the $i^{\mathrm{th}}$ spin. The eigenvalues $\pm 1$ correspond to the spin-up $\vert 0 \rangle$ and spin-down $\vert 1 \rangle$ basis states. This Hamiltonian is equivalent to applying a uniform magnetic field along the $z$-axis, establishing the internal energy structure of the battery via Zeeman splitting \cite{aliqb}.

The initial state of the system is defined as 
\begin{equation}
    \vert\psi(0)\rangle = \vert\downarrow\rangle^{\otimes n}.
    \label{initial state}
\end{equation} 
which is the ground state of the battery Hamiltonian $H_B$ and therefore represents the uncharged configuration of the battery with energy $-n\hbar\omega_0$. The fully charged state, $\vert\uparrow\rangle^{\otimes n}$, has energy $+n\hbar\omega_0$. Thus, the maximum storable or extractable energy in a charging–discharging cycle is $2n\hbar\omega_0$. We represent $\vert\downarrow\rangle$ as $\vert1\rangle$ and $\vert\uparrow\rangle$ as $\vert0\rangle$ in our simulation. \par


To drive the charging process, we introduce the charging Hamiltonian
\begin{equation}
    H_C = H_x + H_{HS} + H_{DMz}.
    \label{charging H}
\end{equation}
which incorporates a transverse magnetic field, Heisenberg interactions, and the Dzyaloshinskii–Moriya interaction (DMI). The first contribution, $H_x$, represents a uniform transverse magnetic field along the $x$-axis
\begin{equation}
    H_{x} = \hbar\Omega\sum_{i = 1}^{n} {\sigma_i^x},
    \label{H x}
\end{equation}
where $\Omega$ is the field strength applied equally to all spins. 
\textcolor{black}{The second term, $H_{HS}$, accounts for the Heisenberg interaction. The one-dimensional Heisenberg Hamiltonian describing spin–spin coupling in the chain is given by \cite{blundell2001magnetism, Ch1_Heisenberg_IOP}
\begin{equation}
H_{HS} = \sum_{<i,j>} J_{ij} \vec{\sigma_i} \cdot \vec{\sigma_j} ,
\label{HS}
\end{equation}
where $J_{ij}$ is the coupling constant between the $i^{\mathrm{th}}$ and $j^{\mathrm{th}}$ spins; in most of our analysis, we take $J_{ij} = J$. Here, $\vec{\sigma_i} = (\sigma_i^x, \sigma_i^y, \sigma_i^z)$ denotes the vector of Pauli operators acting on the $i^{\mathrm{th}}$ spin state.} As shown in Ref.~\cite{many-body-QB}, a spin chain with isotropic coupling evolves much like a collection of independent spins, providing little scope for collective enhancement. To enable cooperative charging, we therefore adopt the anisotropic antiferromagnetic Heisenberg Hamiltonian
\begin{equation}
    H_{HS} = J \sum_{i<j} \Big[ (1+\delta) \sigma_i^x \sigma_{j}^x + (1-\delta) \sigma_i^y \sigma_{j}^y + \Delta \sigma_i^z \sigma_{j}^z \Big]
    \label{HS H}
\end{equation}
where $J=1$ is the nearest-neighbor coupling constant, while $\delta$ and $\Delta$ control anisotropy in the $xy$-plane and along the $z$-direction respectively. 
Based on the values of the anisotropy parameters, we focus on two representative cases of the Heisenberg Hamiltonian
\begin{enumerate}
    \item Ising model: $\delta = 1$, $\Delta = 0$,  
    \item XXZ model: $\delta = 0$, $\Delta \neq 0$.
\end{enumerate}
\textcolor{black}{The third term in eq. \ref{charging H} accounts for the Dzyaloshinskii–Moriya interaction (DMI) \cite{DM1, DM2, DM3, yang2023first}, an antisymmetric and anisotropic exchange interaction that originates from spin–orbit coupling \cite{Sahu_2017, TCHOFFO2016358}. The general form of the DMI Hamiltonian is}

\begin{equation}
H_{DM} = \sum_{i<j} \vec{D}\cdot(\vec{\sigma}_i \times \vec{\sigma}_j),
\label{DM H}
\end{equation}

\textcolor{black}{where $\vec{D}=(D_x, D_y, D_z)$ is the DMI vector, and $D_i$ denotes its strength along the $i^{\mathrm{th}}$ spatial direction with $i \in \{x, y, z\}$. In this work, we restrict our attention to the $z$-component of the DMI, leading to}

\begin{equation}
\textcolor{black}{H_{DMz} = D_z \sum_{i<j} \left( \sigma_i^x\sigma_{j}^y - \sigma_i^y\sigma_{j}^x \right),}
\label{DM z}
\end{equation}

\textcolor{black}{where $D_z$ quantifies the interaction strength along the $z$-axis.}

\textcolor{black}{Experimental and first-principles studies indicate that DMI strengths can vary significantly across materials. For instance, in two-dimensional ternary compounds of the form $MCuX_2$ (with $M$ a $3d$ transition metal and $X$ a group-VIA element), the DMI can range from 0 to 15 meV/atom, with $VCuTe_2$ exhibiting a value of 15.2 meV/atom \cite{Ga_DMI}. Likewise, in 2D Janus structures such as Fe/Ir(111) thin films, a DMI strength of approximately 1.7 meV has been reported, contributing to skyrmion formation \cite{Dupe_Fe_IR}.}

\textcolor{black}{Within the quantum-battery setting, the Heisenberg interaction governs the intrinsic spin–spin coupling responsible for cooperative charging, while the inclusion of DMI introduces anisotropy and spin canting, thereby modifying the population dynamics, enhancing quantum correlations \cite{Gurkan2010943, Jafari20111057}, and altering the available pathways for energy transfer. These combined effects influence both ergotropy and charging power \cite{Rahman2024}, making Heisenberg exchange and DMI crucial components in understanding and optimizing the performance of spin-chain-based quantum batteries.}

In the charging protocol, the quantum system is initialized in the ground state of the battery Hamiltonian $H_B$. At time $t=0$, the charging process is initiated by abruptly switching to the charging Hamiltonian $H_C$. After a time interval $\tau$, the Hamiltonian is switched back to $H_B$, thereby completing the charging cycle. To investigate the role of the battery Hamiltonian during the charging dynamics, we introduce a weighted contribution of $H_B$ to the charging Hamiltonian \cite{shukla2025}. The time-dependent Hamiltonian is thus defined as
\begin{equation}
    H(t) = 
    \begin{cases}
        H_B, & t<0,\\
        H_{C} + \lambda H_B, & 0\le t\le\tau,\\
        H_B, & t>\tau,
    \end{cases}
    \label{driving H}
\end{equation}
where $\lambda \in [0,1]$ controls the contribution of the battery Hamiltonian during the charging interval. For $\lambda=0$, the dynamics is governed purely by the charging Hamiltonian $H_C$, while for $\lambda>0$ the battery Hamiltonian partially contributes during charging. \par
In this sense, $\lambda$ acts as a continuous switch that interpolates between purely driven dynamics ($\lambda=0$) and dynamics where the battery Hamiltonian partially contributes during charging ($\lambda>0$). This goes beyond a simple ON/OFF protocol and allows us to probe intermediate regimes that are commonly relevant in realistic experimental settings. \par
Since the Hamiltonian is static during the charging window $0 \le t \le \tau$, the unitary time evolution operator governing the charging dynamics is given by
\begin{equation}
    U = e^{-i (H_C + \lambda H_B)t}.
\end{equation}
Throughout the charging process, the system is assumed to be thermally isolated, and all energy stored in the battery is quantified with respect to the battery Hamiltonian $H_B$, which defines the reference energy spectrum used in the computation of ergotropy.\par

These Hamiltonians establish the theoretical framework for the charging dynamics of our quantum battery. In the next section, we examine how different geometrical configurations of the spin chain, namely open boundary, closed boundary, and supercube topologies, affect the storage and extraction of energy.

\begin{figure*}[ht]
    \centering
    \includegraphics[width=\linewidth]{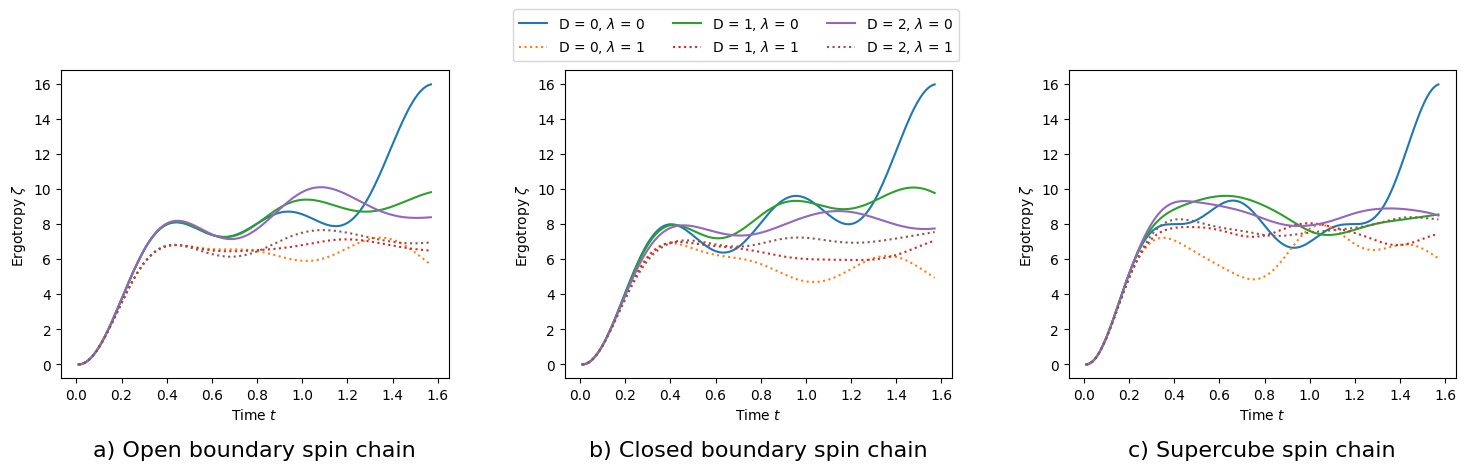}
    \caption{Ergotropy for Ising model ($\delta = 1, \Delta = 0$) with DM interaction strength D = 0, 1, 2}
    \label{fig:diff d erg ising}
\end{figure*}

\begin{figure*}[ht]
    \centering
    \subfigure[Open boundary spin chain]{\includegraphics[width = 0.3\linewidth]{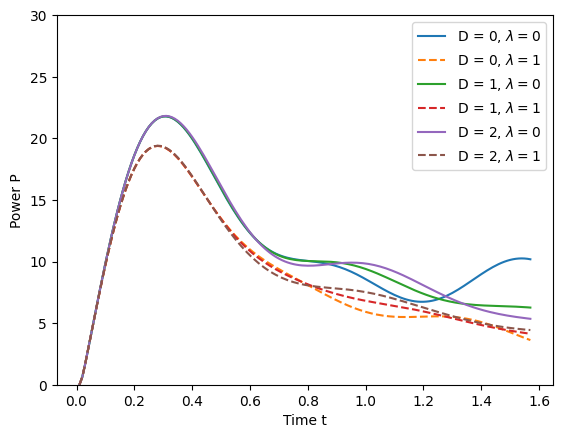}}
    \subfigure[Closed boundary spin chain]{\includegraphics[width = 0.3\linewidth]{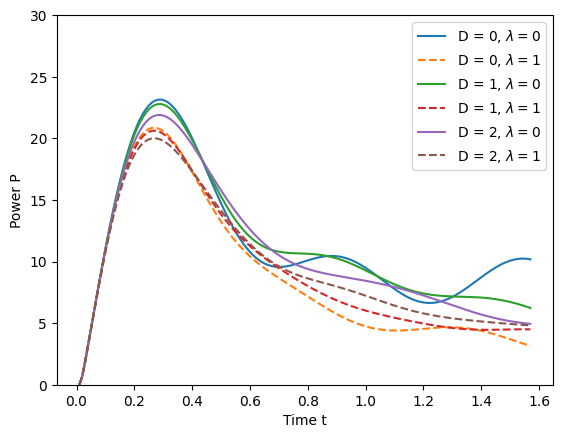}}
    \subfigure[Supercube spin chain]{\includegraphics[width = 0.3\linewidth]{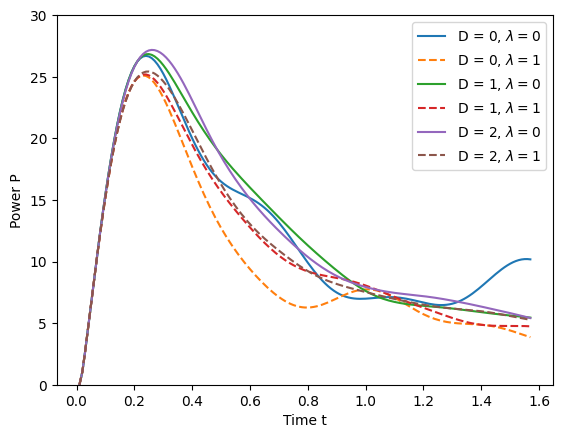}}
    \caption{Power for Ising model ($\delta = 1, \Delta = 0$) with DM interaction strength D = 0, 1, 2}
    \label{fig:diff d power ising}
\end{figure*}

\subsection{Spin chain configurations \label{sec:3}}
We consider three distinct spin chain configurations for each Hamiltonian model. In all cases, the interaction Hamiltonian $H_{int}$ is defined as the sum of the anisotropic Heisenberg interaction term $H_{HS}$ and the Dzyaloshinskii–Moriya (DM) interaction term $H_{DMz}$. 


\subsubsection{Open boundary spin chain configuration}
In the open boundary configuration, the spins are arranged linearly under open boundary conditions, such that each spin interacts only with its nearest neighbor \cite{Zheng2025}. The interaction Hamiltonian is given by
\begin{equation}
    \begin{aligned}
    H_{int} &= J \sum_{i = 1}^{n - 1} \Big[ (1+\delta) \sigma_i^x \sigma_{i+1}^x + (1-\delta) \sigma_i^y \sigma_{i+1}^y + \Delta \sigma_i^z \sigma_{i+1}^z \Big]\\ 
    &+ D \sum_{i = 1}^{n - 1} \Big( \sigma_i^x \sigma_{i+1}^y - \sigma_i^y \sigma_{i+1}^x \Big).
\end{aligned}
\label{interaction H open}
\end{equation}

\subsubsection{Closed boundary spin chain configuration}

In the closed boundary configuration, the last spin interacts with the first spin, forming a closed loop. This implements periodic boundary conditions, ensuring that each spin has two neighbors, even at the boundaries \cite{Zheng2025}. The interaction Hamiltonian is therefore similar to the open boundary case but extended to include the periodic connection between the first and last spins such that
\begin{equation}
    \begin{aligned}
    H_{int} &= J \sum_{i = 1}^{n} \Big[ (1+\delta) \sigma_i^x \sigma_{i+1}^x + (1-\delta) \sigma_i^y \sigma_{i+1}^y + \Delta \sigma_i^z \sigma_{i+1}^z \Big]\\ 
    &+ D \sum_{i = 1}^{n} \Big( \sigma_i^x \sigma_{i+1}^y - \sigma_i^y \sigma_{i+1}^x \Big),
\end{aligned}
\label{interaction H closed}
\end{equation}
where $\sigma_{n+1}^j = \sigma_1^j$ and $j \in \{x, y, x\}$

\subsubsection{Supercube spin chain configuration}
The supercube configuration is inspired by the connectivity structure of quantum neural networks \cite{QNN}. In this model, the system qubits are arranged in a one-dimensional array, but their interactions are mapped onto the structure of a cube, where the couplings between qubits are represented as edges. To further enrich the connectivity, additional couplings are introduced along the face diagonals and body diagonals. This extended interaction network allows us to study the effect of higher connectivity on the charging performance of the quantum battery. Consequently, the interaction Hamiltonian becomes more complex, incorporating nearest-neighbor, next-nearest-neighbor, and longer-range couplings, each accompanied by corresponding Dzyaloshinskii–Moriya (DM) interaction terms. The corresponding interaction Hamiltonian is

\begin{equation}
    \begin{aligned}
    H_{int} &= \Big\{ J \sum_{i = 1, i\in Z_{odd}}^{8} [ (1+\delta) \sigma_i^x \sigma_{i+1}^x + (1-\delta) \sigma_i^y \sigma_{i+1}^y\\ &+ \Delta \sigma_i^z \sigma_{i+1}^z ] 
    + D \sum_{i = 1, i\in Z_{odd}}^{8} ( \sigma_i^x \sigma_{i+1}^y - \sigma_i^y \sigma_{i+1}^x ) \Big\}\\
    &+ \Big\{ J \sum_{i = 1, 2, 5, 6} [ (1+\delta) \sigma_i^x \sigma_{i+2}^x + (1-\delta) \sigma_i^y \sigma_{i+2}^y\\ &+ \Delta \sigma_i^z \sigma_{i+2}^z ]
    + D \sum_{i = 1, 2, 5, 6} ( \sigma_i^x \sigma_{i+2}^y - \sigma_i^y \sigma_{i+2}^x ) \Big\}\\
    &+ \Big\{ J \sum_{i = 1}^{4} [ (1+\delta) \sigma_i^x \sigma_{i+4}^x + (1-\delta) \sigma_i^y \sigma_{i+4}^y + \Delta \sigma_i^z \sigma_{i+4}^z ]\\ 
    &+ D \sum_{i = 1}^{4} ( \sigma_i^x \sigma_{i+4}^y - \sigma_i^y \sigma_{i+4}^x ) \Big\}.
\end{aligned}
\label{interaction H supercube}
\end{equation}

\begin{figure*}[ht]
    \centering
    \includegraphics[width=\linewidth]{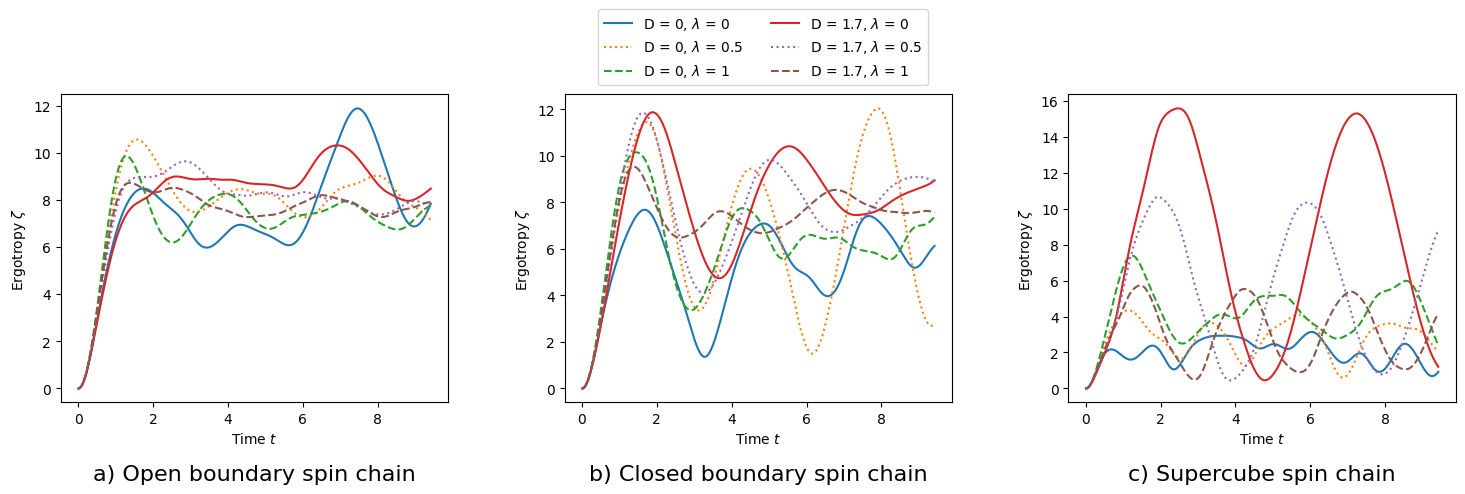}
    \caption{Ergotropy for XXZ model with DM interaction strength D = 0, 1.7}
    \label{fig:erg XXZ}
\end{figure*}

\begin{figure*}[ht]
    \centering
    \subfigure[Open boundary spin chain]{\includegraphics[width = 0.3\linewidth]{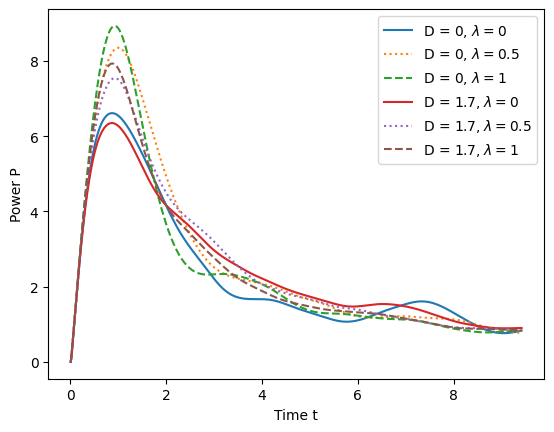}}
    \subfigure[Closed boundary spin chain]{\includegraphics[width = 0.3\linewidth]{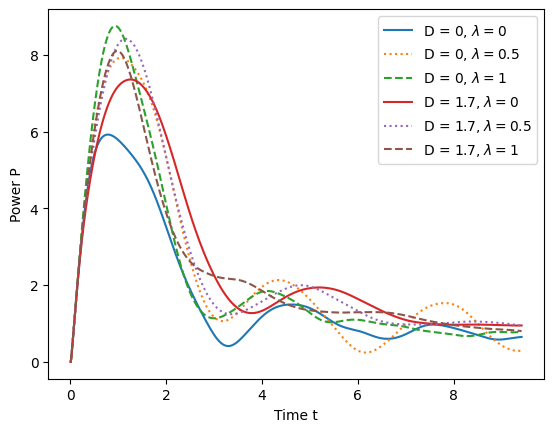}}
    \subfigure[Supercube spin chain]{\includegraphics[width = 0.3\linewidth]{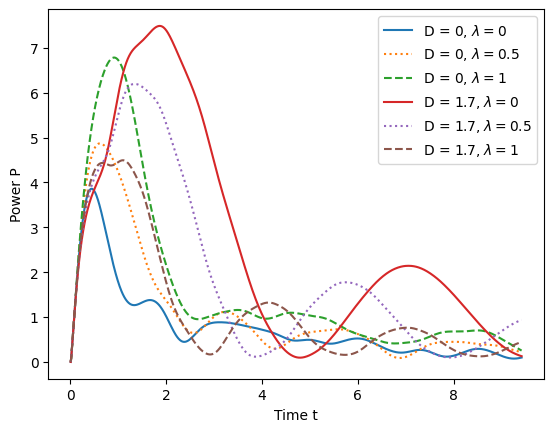}}
    \caption{Power for XXZ model with DM interaction strength D = 0, 1.7}
    \label{fig:power XXZ}
\end{figure*}
These Hamiltonians establish the model configurations of interest. To assess their suitability as quantum batteries, we now introduce the performance metrics used in this study.

\begin{figure}[ht]
    \centering
    \subfigure[Ergotropy with varying DM interaction strength ($J=1$) \label{fig-a:erg vary DM}]{\includegraphics[width = 0.65\linewidth]{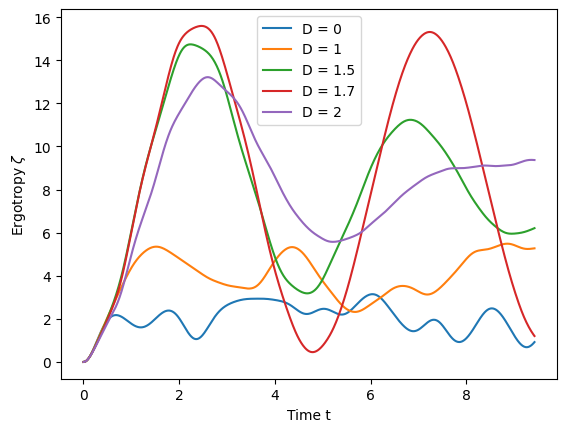}}
    \subfigure[Ergotropy with varying Heisenberg interaction strength ($D=1.7$) \label{fig-b:erg vary J}]{\includegraphics[width = 0.65\linewidth]{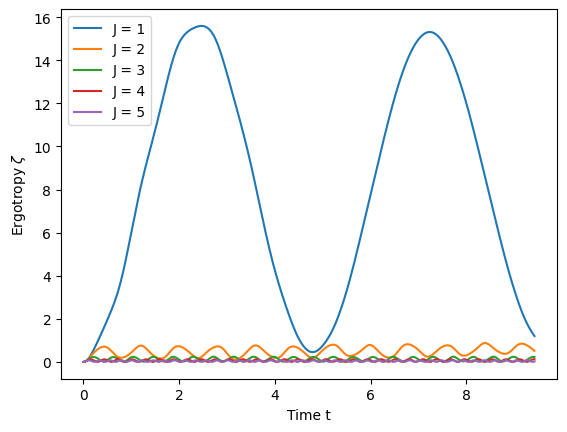}}
    \caption{Ergotropy and Power for supercube spin chain configuration in XXZ model with varying interaction strengths}
    \label{fig:erg p cube}
\end{figure}

\subsection{QB Performance Metrics \label{sec:3b}}

To evaluate the performance of the spin chain quantum batteries, we focus on two standard metrics: ergotropy and charging power.

\subsubsection{Ergotropy}
Ergotropy is defined as the maximum amount of work that can be extracted from a quantum battery \cite{Allahverdyan_2004, Sone2021, Tirone2021, Touil_2022}. The concept relies on the notion of a passive state, from which no work can be extracted under unitary operations \cite{Pusz1978, Lenard1978, Salvia2021distributionofmean, Mazzoncini2023}. \par

If the battery Hamiltonian can be written as $H_B = \sum_{i}{\epsilon_i\vert\epsilon_i\rangle\langle\epsilon_i\vert}$ where $\epsilon_i$ represents the eigenenergies of $H_B$ and $|\epsilon_i\rangle$ are the corresponding eigenstates ordered by $\epsilon_i<\epsilon_{i+1}$, and the density matrix of the system be $\rho(t)=\sum_{i}{r_i\vert r_i\rangle\langle r_i\vert}$ such that $r_i \ge r_{i+1}$ where $r_i$ represent the eigenvalues of $\rho(t)$ and $|r_i\rangle$ are the corresponding eigenvectors, then passive state of the system can be defined as $\rho_p = \sum_{i}{r_i}\vert\epsilon_i\rangle\langle\epsilon_i\vert$ \cite{Francica2020}. Thus ergotropy will be

\begin{equation}
    \zeta = \text{Tr}[H_B(\rho(t)-\rho_p)].
\end{equation}

For pure states, the ground state of the system is passive ($\rho_p = {r_0}\vert\epsilon_0\rangle\langle\epsilon_0\vert$, $r_0=1$) \cite{Passive_state}. Accordingly, the ergotropy is given by the difference in energy between the evolved state $\vert\psi(t)\rangle = U\vert\psi(0)\rangle$ and the ground state $\vert\psi(0)\rangle$, i.e.
\begin{equation}
    \zeta = \langle\psi(t)\vert H_B \vert\psi(t)\rangle - \epsilon_0,
    \label{ergotropy}
\end{equation}
where, $\vert \psi (0) \rangle = (|\downarrow\rangle)^{\otimes n}$ is the initial state of the system, which is also the ground state of $H_B$, and $\epsilon_0$ denotes passive state energy, i.e., the lowest eigenvalue of $H_B$ given by $\langle\psi(0)|H_B|\psi(0)\rangle$. $H_{B}$ denotes the internal Hamiltonian of the battery as defined in Eq. (5).

\begin{figure*}[ht!]
    \centering
    \subfigure[Adding two cross-body diagonals \label{fig-a:body diag}]{\includegraphics[width = 0.35\linewidth]{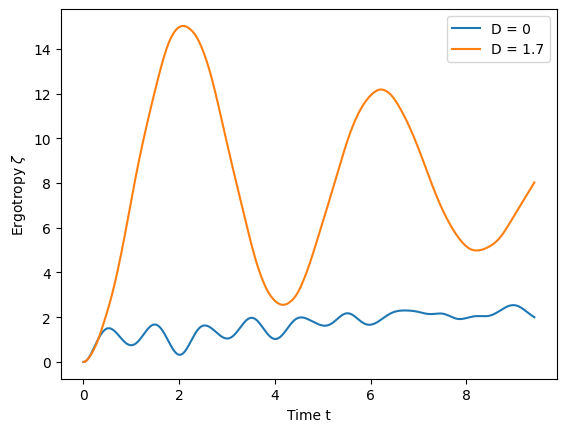}}
    \subfigure[Adding top face diagonals\label{fig-b:face diag}]{\includegraphics[width = 0.35\linewidth]{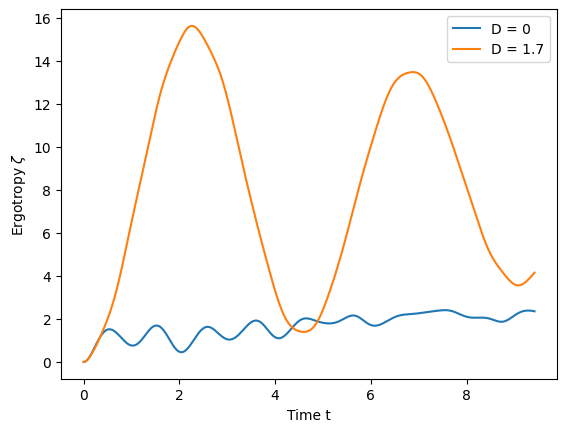}}
    \caption{Adding connections corresponding to body diagonal and face diagonal}
    \label{fig:add cube}
\end{figure*}

\subsubsection{Charging Power}

Charging power quantifies the rate at which extractable work is stored in the system during the charging process. It is defined as
\begin{equation}
    P(t) = \frac{\zeta}{t}.
    \label{power}
\end{equation}
A higher charging power indicates faster charging and is therefore desirable for practical applications.

\section{Results \label{sec:4}}
In this section we present the results for the eight-qubit quantum battery under different spin-chain configurations and Hamiltonian models. Our analysis focuses on the Ising and XXZ models, and compares their performance in terms of ergotropy and charging power. We also examine the influence of the battery Hamiltonian $H_B$ on the charging process by varying the parameter $\lambda \in [0, 1]$. In addition, we extend our analysis of QB architecture in XXZ model to varying system sizes, to explore how number of qubits influence charging performance. \textcolor{black}{The population dynamics plots for all three geometries in Ising model and XXZ model for $D=0, 1.7$ when $\lambda = 0$ are included in the Appendix \ref{A1}. While the main text discusses the qualitative trends and their implications for energy transfer and ergotropy, the detailed population evolution offers an additional layer of transparency regarding how individual spin levels redistribute during the charging process. These plots help verify that the observed charging behavior is not an artefact of specific initial conditions or restricted parameter choices, and they serve as a diagnostic tool to distinguish whether features such as enhanced charging arise from interaction-driven dynamics rather than entanglement effects. Including these results in the Appendix allows one to examine the full dynamical behavior without interrupting the flow of the main discussion, while still ensuring that all supporting analyzes remain accessible.} \par
Unless stated otherwise, the values of the parameters used in this study are as follows: number of qubits $n = 8$, reduced Planck constant $\hbar = 1$, Larmor frequency $\omega_0 = 1$, transverse field strength $\Omega = 1$, and Heisenberg coupling constant $J = 1$. For the Ising model, the anisotropy parameters are $\delta = 1, \Delta = 0$, corresponding to the standard Ising model, and the DMI strength is varied as $D \in \{0, 1, 2\}$ to examine the effect of increasing interaction strength. For the XXZ model, we set $\delta = 0, \Delta = 2$, representing anisotropy along the $z$-axis, and consider $D \in \{0, 1.7\}$, where $D=1.7$ is chosen as a representative moderate value to illustrate the impact of finite DMI without reaching the saturation regime.
In the following subsections, we compare the results for each Hamiltonian model across the different spin-chain configurations.

\subsection{Ising model}

\subsubsection{Without DM interaction}

We first analyze the Ising model without DMI, as shown in Figs.~\ref{fig:erg ising} and \ref{fig:power ising}. \textcolor{black}{We study the effect of the battery Hamiltonian $H_B$ on the ergotropy of the system and compare the charging power across the three configurations, evolving each system until it reaches its maximum ergotropy value $= 16$ (i.e., at $t = \pi/2$).} In Fig.~\ref{fig:erg ising}, the ergotropy of all three spin-chain configurations reaches the maximum value $2n\hbar\omega_0 = 16$ at $t=\pi/2$ for $\lambda=0$. This corresponds to full charging under the action of the charging Hamiltonian $H_C$ alone. As $\lambda$ increases, the maximum extractable energy decreases (orange curve, $\lambda=0.5$; green curve, $\lambda=1$), showing that the inclusion of $H_B$ suppresses energy storage across all configurations. \par
Figure~\ref{fig:power ising} compares the charging power. The supercube configuration shows the highest peak power, followed by the closed chain and then the open chain. The peaks occur at $t < \pi/4$, indicating fast charging in all cases. For $\lambda > 0$, the charging power decreases, with the divergence from the $\lambda=0$ curve becoming more pronounced at later times. This confirms that $H_B$ is detrimental both to the maximum stored energy and to the charging rate. Notably, the supercube retains relatively higher power even when $\lambda > 0$, likely due to its richer interaction geometry.

\subsubsection{With DMI}
We next examine the effect of including DMI in the Ising model \textcolor{black}{and analyze the changes in peak power and ergotropy curves for $D=0, 1, 2$ when the system is evolved until ergotropy attains its maximum value of $16$ for $D=0$}, shown in Figs.~\ref{fig:diff d erg ising} and \ref{fig:diff d power ising}. Introducing DMI leads to a significant reduction in peak ergotropy compared with the $D=0$ case. For any $D$, increasing $\lambda$ further reduces the peak ergotropy, again reflecting the suppressive role of $H_B$. \par
\textcolor{black}{The effect on charging power depends on interaction geometry. For supercube chain, larger $D$ enhances peak power, indicating higher charging rates. While in open boundary spin chain, increasing $D$ from 0 to 2 does not show any significant improvement in maximum charging power. The three curves ($D= 0, 1, 2$) overlap until $t\approx\pi/4$ and diverge slightly at later times. Remarkably, when $D$ is increased beyond 2, peak power for the system also increases.} By contrast, in the closed chain, increasing $D$ reduces the maximum power. For all cases, including $H_B$ ($\lambda > 0$) visibly suppresses the peak power. Moreover, when $D$ greatly exceeds the Ising coupling $J$, the maximum power saturates, showing diminishing returns for very strong DMI \cite{zhang2024quantum}. \par
Overall, the supercube configuration consistently outperforms the open and closed chains, maintaining higher charging power across all values of $D$ and $\lambda$. We now turn to the XXZ model to examine how anisotropy along the $z$-axis alters the charging dynamics.

\subsection{XXZ model}
As seen in the previous section, increasing DMI strength enhances the performance of the QB in the Ising model for supercube configuration, while exhibiting a negative effect for the closed boundary configuration. We now study the effect of DMI on the three spin configurations of the QB in the XXZ model. We analyze ergotropy and charging power for $D = 0$ and $1.7$, and also examine the influence of the battery Hamiltonian $H_B$ by varying the parameter $\lambda$, as shown in Figs.~\ref{fig:erg XXZ} and \ref{fig:power XXZ}. The plots are presented for $t \in [0, 3\pi]$ as it captures at least one entire oscillation cycle. \textcolor{black}{For reference, the maximum ergotropy for our 8 qubit system is 16. And our focus is on the periodicity of ergotropy curve as it indicates the ability of the system to fully charge and then reset to 0.} Results are discussed separately for each spin-chain configuration.

\subsubsection{Open boundary spin chain configuration}
From the ergotropy plot in Fig.~\ref{fig:erg XXZ}(a), we observe that the curves lack regular oscillatory behavior. The system never reaches its fully charged state, hence ergotropy is always less than the maximum value of $16$. Even upon introducing DMI and the battery Hamiltonian into the system, the value of ergotropy fluctuates within a certain range without reaching higher amplitudes. \par

Figure~\ref{fig:power XXZ}(a) shows the corresponding charging power. Consistent with the ergotropy results, DMI provides little advantage. For intermediate times $t \gtrsim 2$ up to $t \approx 6$, there is a slight increase in power for $D=1.7, \lambda=0$, but no significant enhancement in stability. For $\lambda>0$, we observe an increase in peak power which increases as $\lambda$ increases, with DMI suppressing the peak value initially and later converging to the $D = 0$ case. Overall, the open chain gains limited benefit from DMI under the XXZ model.

\subsubsection{Closed boundary spin chain configuration}
In the closed boundary case [Fig.~\ref{fig:erg XXZ}(b)], introducing DMI ($D=1.7$) leads to a marked increase in peak ergotropy, accompanied by sharp oscillations. However, the system discharges asymmetrically, i.e., after each discharge it does not return fully to the passive state but remains partially charged, leaving residual energy that cannot be extracted. This reduces cycle efficiency. Notably, the minima of the oscillations show an upward drift, indicating energy accumulation over repeated cycles. When $\lambda = 0.5$, peak ergotropy increases but when $\lambda = 1$, there is a drop in peak value. For $D=1.7$, ergotropy remains consistently higher than the $D=0$ case for $\lambda = 0, 1$, but for $\lambda = 0.5$ both the curves initially follow the same trend, but diverge significantly afterwards. \par

Figure~\ref{fig:power XXZ}(b) shows that DMI significantly enhances the peak power for $\lambda = 0$ and similarly for $\lambda = 0.5$ DMI results in a slight increase in peak value. However, for $\lambda = 1$, DMI suppresses maximum power. Increasing $\lambda$ increases the increases the maximum power for $D=0$ case, but a slight decrease is observed for $D = 1.7$ when $\lambda = 1$. Thus, the closed boundary configuration benefits from DMI in the XXZ model, with the drawback of accumulating residual energy across cycles.

\begin{figure*}[ht!]
    \centering
    \subfigure[$4$ qubits \label{fig: 4 qubits graph}]{\includegraphics[width = 0.25\linewidth]{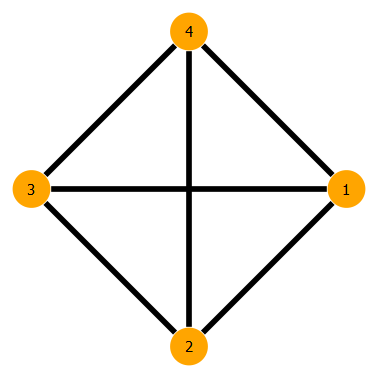}}\hfill
    \subfigure[$6$ qubits \label{fig: 6 qubits graph}]{\includegraphics[width = 0.25\linewidth]{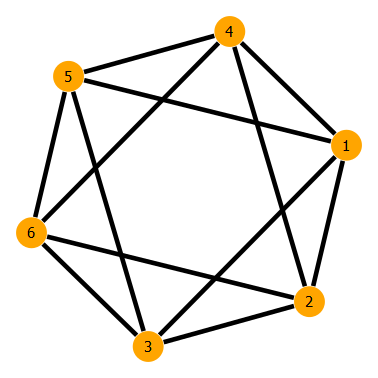}}\hfill
    \subfigure[Icosahedron \label{fig: icosahedron}]{\includegraphics[width = 0.25\linewidth]{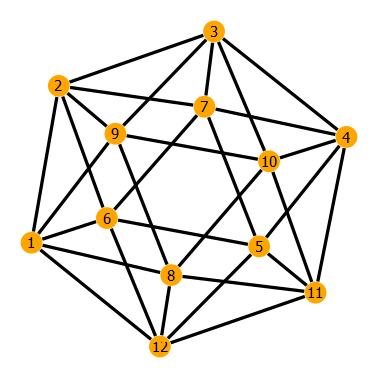}}\hfill 
    \subfigure[\label{4 qubits ergotropy}]{\includegraphics[width = 0.3\linewidth]{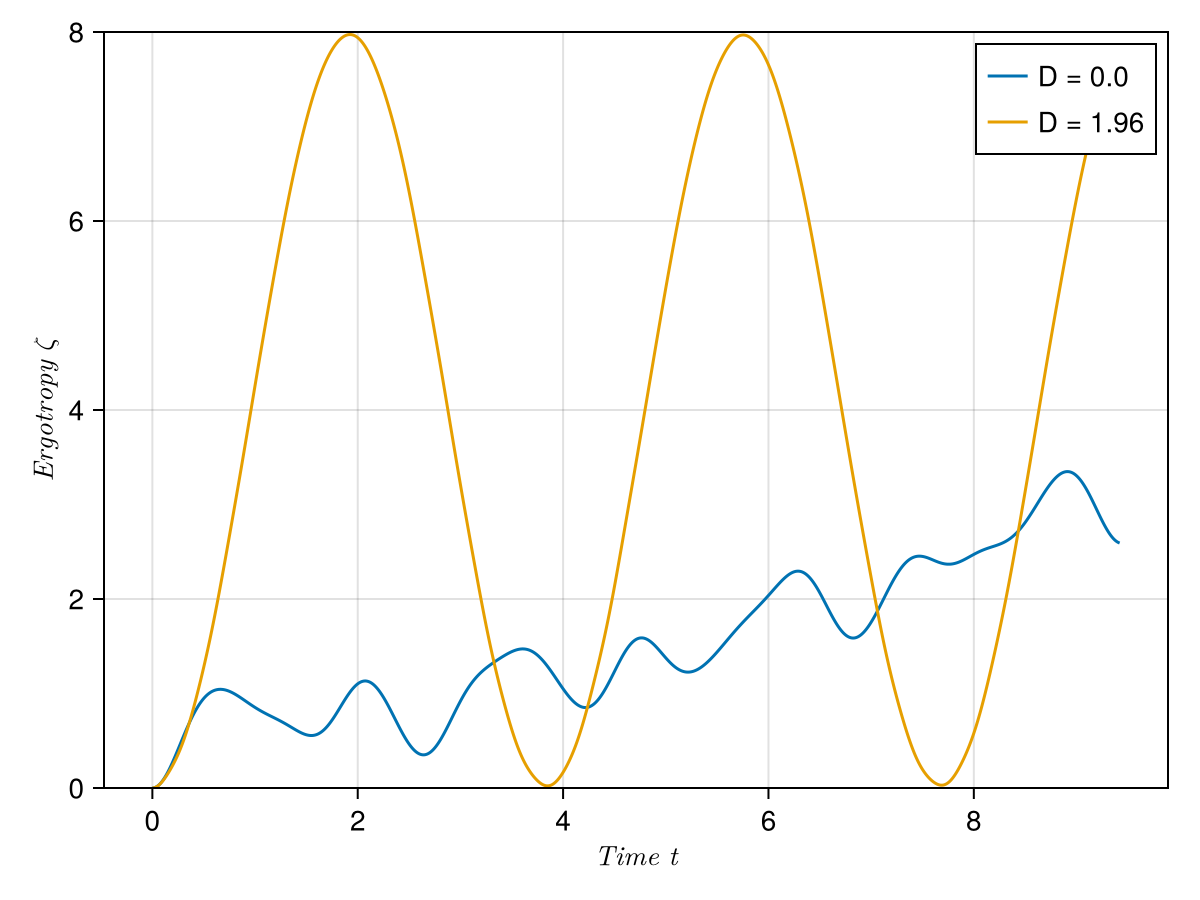}}
    \subfigure[\label{6 qubits ergotropy}]{\includegraphics[width = 0.3\linewidth]{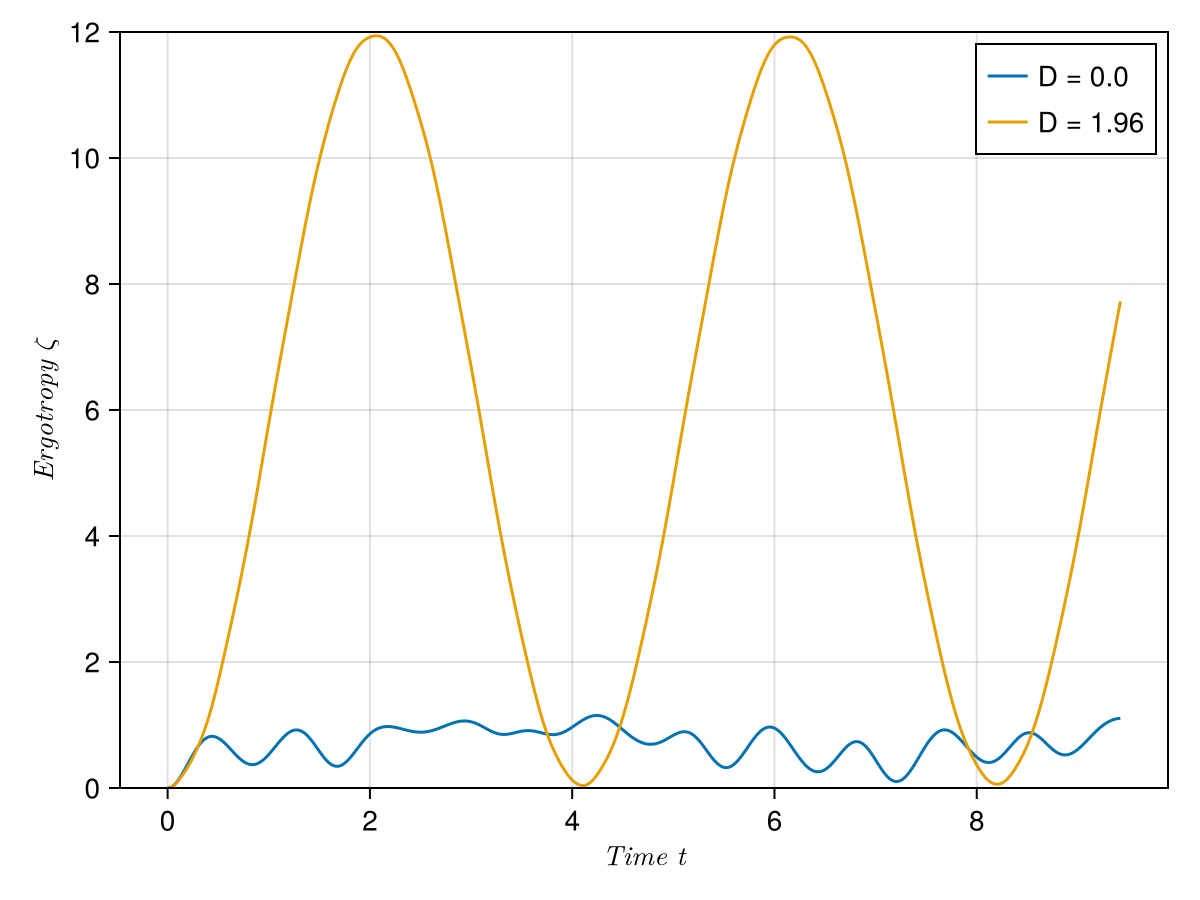}}
    \subfigure[\label{fig: icosa erg}]{\includegraphics[width = 0.3\linewidth]{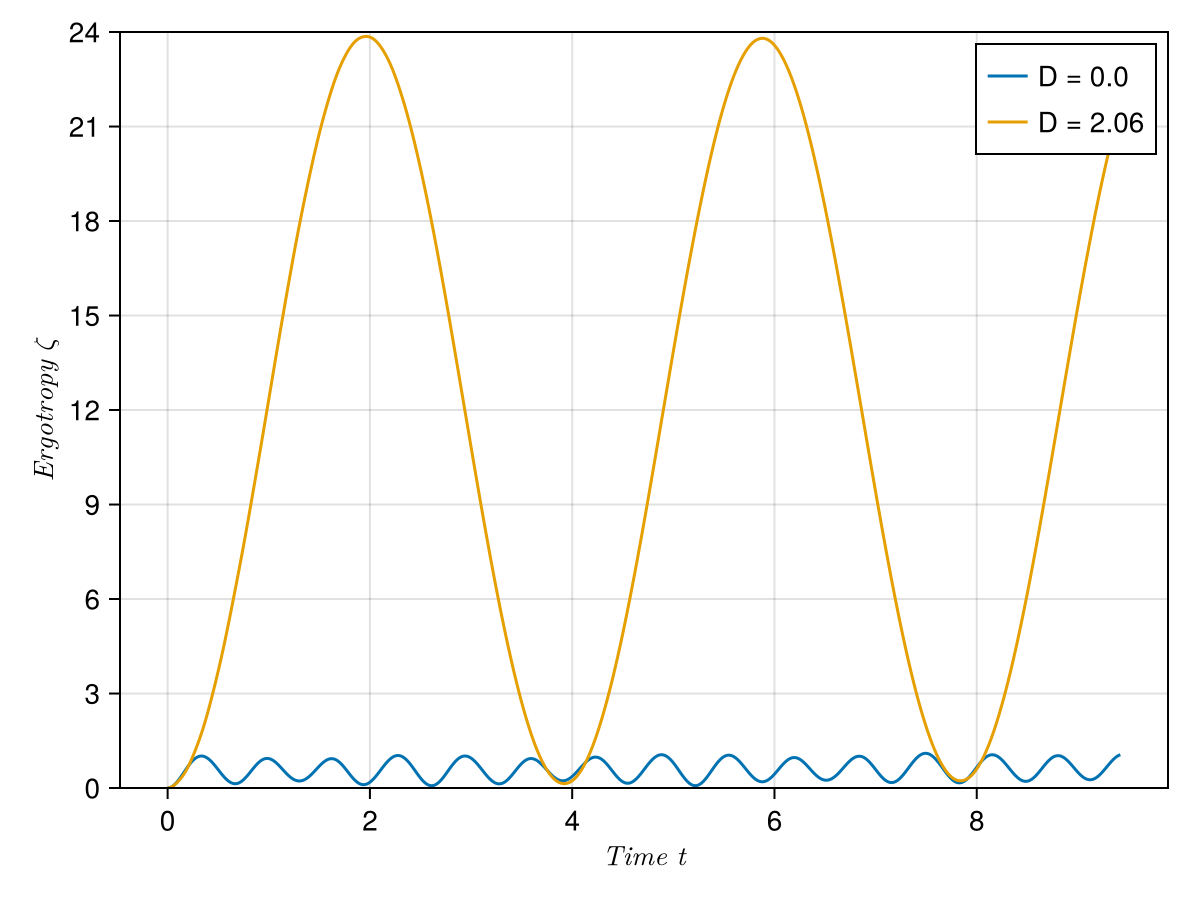}}\hfill
    \caption{Structures $(a-c)$ and Ergotropy time evolution $(d-f)$ for symmetric Platonic geometries: (a) tetrahedron $(n=4)$, (b) octahedron $(n=6)$, and (c) icosahedron $(n=12)$. All cases exhibit perfect sinusoidal charging–discharging cycles at their respective optimal DMI strengths, demonstrating that highly symmetric connectivities preserve ideal QB operation across different system sizes.}
    \label{fig:4q6q12q}
\end{figure*}

\subsubsection{Supercube spin chain configuration}
Figures~\ref{fig:erg XXZ}(c) and \ref{fig:power XXZ}(c) show that the supercube configuration demonstrates the most desirable features for a QB under the XXZ model. The ergotropy curve with DMI exhibits a clean sinusoidal charging–discharging cycle, periodically reaching close to the maximum value and returning to the passive state with negligible residual energy. This reset capability makes it well suited for rechargeable QB models and clocked quantum protocols. The sinusoidal behavior remains stable over extended times, reflecting robust energy exchange between the battery and charging field facilitated by DMI \cite{energy_transfer}. For $\lambda>0$ at $D=1.7$, the cycle persists, though peak ergotropy decreases with increasing $\lambda$. Importantly, ergotropy remains well above the $D=0$ case throughout except when $\lambda = 1$, showing that inclusion of the battery Hamiltonian plays a detrimental role in the charging dynamics.\par

Looking at the power curves shown in figure \ref{fig:power XXZ}(c), we note that supercube exhibits the highest peak among all three configurations for $D=1.7$ when $\lambda=0$, slightly surpassing that for closed boundary spin chain. Power eventually decays over time but DMI leads to broader power curves except when the battery Hamiltonian is included in the charging dynamics. Overall, among the three configurations, the supercube consistently outperforms both open and closed chains in the XXZ model for $\lambda = 0$, mirroring its advantage in the Ising case. \par
Across all the Hamiltonian models, the supercube configuration consistently delivers superior performance, combining high ergotropy, strong charging power, and stable charging cycles. In the Ising model, DMI enhances charging power for open and supercube chains but degrades performance in the closed chain. In contrast, under the XXZ model, DMI significantly improves the closed and supercube chains, with the latter displaying clean sinusoidal cycles and negligible residual energy, making it particularly attractive for rechargeable QB applications. Thus, while the influence of DMI depends on both model and topology, the supercube architecture remains the most robust and efficient design. Motivated by the superior performance of the supercube configuration across both Ising and XXZ models, we now examine how variations in interaction parameters and structural modifications affect its charging behavior.

\paragraph{Varying interaction strengths in Supercube:}

We study the effect of varying the DMI strength $D$ and the Heisenberg coupling constant $J$ on the charging behavior of the supercube spin chain in XXZ model illustrated in figure \ref{fig:erg p cube}. Figure \ref{fig-a:erg vary DM} shows ergotropy for different values of $D$ keeping $J = 1$ fixed. When $D \neq 1.7$, the optimal sinusoidal charging-discharging cycle disappears and the maximum ergotropy decreases, indicating that there exists an optimal DMI strength for sustaining high-efficiency periodic energy transfer between the battery and the charging field. Figure \ref{fig-b:erg vary J} shows ergotropy for different values of $J$ keeping $D = 1.7$ fixed. The periodic charging cycle disappears for $J \neq 1$, and ergotropy of the system significantly diminishes, becoming nearly zero for $J>2$. These observations suggest that the combination $D=1.7$ and $J=1$ are optimal points for the supercube QB model under XXZ Hamiltonian. This behavior further indicates the presence of a resonance-like condition, where the balance between exchange interaction ($J$) and spin–orbit coupling effects induced by DMI enables efficient cyclic energy transfer. \par

\paragraph{Adding diagonals in supercube configuration:}

We investigate the effect of adding additional connections between the qubits in supercube configuration. In figure \ref{fig-a:body diag}, we introduce interactions between the qubits located at the end points of two cross-body diagonals of the supercube. In figure \ref{fig-b:face diag} we add interactions along the face diagonals on the top face of the cube. In both cases, we observed a noticeable decay in the amplitude of the charging-discharging cycle. When all four cross-body diagonals are added, the oscillation amplitude decreases further. Adding all face diagonals results in complete suppression of the sinusoidal nature of charging curve. Similarly, the sinusoidal nature vanishes when both body and face diagonals are introduced simultaneously into the system. These observations suggest that the addition of connections between qubits beyond the original supercube structure disrupts the energy exchange that supports optimal battery performance, and plays a detrimental role in charging of the supercube configuration. Hence, the original supercube model appears to represent a finely tuned structure to facilitate efficient charging cycles. \par

\begin{table*}[ht]
\begin{tabular}{|c|c|c|c|} 
 \hline
 No. of qubits (n) & No. of connections (c) & DMI strength (D) \\ [0.5ex] 
 \hline\hline
 3 & 2 & 1.86 \\
 \hline
 4 & 3 & 1.96 \\ 
 \hline
 5 & 4 & 2.03 \\
 \hline
 \multirow{2}{0.5em}{6} & 4 & 2.18 \\
 & 5 & 2.07 \\
 \hline
 7 & 6 & 2.11 \\
 \hline
 \multirow{3}{0.5em}{8} & 5 & 2.30 \\
 & 6 & 2.23 \\
 & 7 & 2.13 \\
 \hline
 \multirow{2}{0.5em}{9} & 6 & 2.36 \\
 & 8 & 2.15 \\
 \hline
 \multirow{3}{0.5em}{10} & 7 & 2.33 \\
 & 8 & 2.25 \\
 & 9 & 2.17 \\
 \hline
 \multirow{2}{0.5em}{11} & 8 & 2.35 \\
 & 10 & 2.18 \\
 \hline
 \multirow{5}{0.5em}{12} & 7 & 2.52 \\
 & 8 & 2.46 \\
 & 9 & 2.33 \\
 & 10 & 2.27 \\
 & 11 & 2.20 \\
 \hline
\end{tabular}
\caption{Different quantum battery structures with their corresponding DMI strength (D) for which ergotropy follows a sinusoidal curve}
\label{Table: geometries}
\end{table*}

\subsubsection{Structures with different number of qubits and coordination number \label{3.b.4}}

While the 8-qubit supercube achieves a near-ideal sinusoidal charging-discharging cycle, it is important to assess whether this behavior persists when the system size and coordination numbers are varied. To this end, we examine a family of $c$-regular graphs with qubit numbers ranging from $n=3$ to $n=12$, where each qubit interacts with exactly $c$ neighbors. This framework allows us to treat a broad range of connectivities on equal footing. For every case studied, we identify an optimal DMI strength for which the ergotopy dynamics exhibit an ideal sinusoidal behavior, periodically reaching the theoretical maximum $2n\hbar\omega_0$ and discharging fully to the passive state with negligible residual energy $\approx 0$. The optimal values of DMI interaction for different choices on $n$ and coordination number $c$ are summarized in Table \ref{Table: geometries}. \par
To illustrate these general trends, Figure \ref{fig:4q6q12q} presents results for three representative Platonic solids embedded in this framework- the tetrahedron $(n=4, c=3)$, the octahedron $(n=6, c=4)$, and the icosahedron $(n=12, c=5)$. The tetrahedral case in Figure \ref{fig: 4 qubits graph} demonstrates that even the smallest highly symmetric geometry supports a perfect sinusoidal ergotropy cycle reaching maximum energy $2n\hbar\omega_0 = 8$ and discharging to its ground state, optimized at $D=1.96$ (Figure \ref{4 qubits ergotropy}). The octahedron case in Figure \ref{fig: 6 qubits graph} confirms that increasing coordination to $c=4$ stabilizes the oscillations, again yielding a full discharge at each cycle at $D=1.96$ (Figure \ref{6 qubits ergotropy}). The maximum energy stored by this system is $2n\hbar\omega_0 = 12$. Motivated by this observation, we generalize our analysis to c-regular graphs \cite{Kim2006, meringer1999fast} constructed using \cite{SciPyProceedings_11}, varying both the number of vertices $n$ and the degree $c$. For each graph, we tune the DMI strength $D$ to identify the value at which the ergotropy exhibits a sinusoidal behavior. The interaction Hamiltonian for such c-regular graphs is given by
 
\begin{equation}
\begin{aligned}
    H_{int} & = J \sum_{(i, j) \in edges}{(1+\delta)\sigma^x_i\sigma^x_j + (1 - \delta)\sigma^y_i\sigma^y_j + \Delta\sigma^z_i\sigma^z_j} \\
    & + D \sum_{(i, j) \in edges}{\sigma^x_i\sigma^y_j - \sigma^y_i\sigma^x_j}
\end{aligned}
\label{H edges}
\end{equation}
where \emph{edges} denote the set of all connected node pairs. A regular graph can be generated from $n$ and $c$ provided $c<n$ and $n \times c$ is even. To construct edges, the $n$ nodes are arranged on a circle. For even $c$, each node $i$ is connected to its $\frac{c}{2}$ nearest neighbors in both clockwise and counter-clockwise directions. For odd $c$, each node is connected to its $(c-1)/2$ nearest neighbors in both directions and additionally to the node opposite it, $(i + \frac{n}{2})$. \par

Among all tested structures, the icosahedron (Figure \ref{fig: icosahedron}) provides the clearest evidence of how symmetry governs efficient operation. Its five-fold coordination enables a highly uniform distribution of interactions, which translates into perfectly sinusoidal charging–discharging cycles with a complete reset to the ground state after each period. Unlike the extended cube or even the cuboctahedron, the icosahedron shows no accumulation of residual energy, ensuring that every cycle begins from the true passive state. Moreover, it reaches the maximum ergotropy value $2n\hbar\omega_0 = 24$ more rapidly than less symmetric 12-qubit structures, highlighting that carefully chosen connectivity can offset the efficiency loss normally associated with scaling. This result highlights that high geometric symmetry can recover the efficiency that is otherwise degraded in larger systems.

Taken together, the results from both $c$-regular graphs and Platonic solids demonstrate a general design principle, i.e., highly symmetric connectivities- whether represented as regular graphs or as polyhedral skeletons- preserve ideal sinusoidal ergotropy cycles across different system sizes. In particular, the icosahedron highlights that scalability to larger qubit numbers is possible without sacrificing performance, provided the underlying connectivity remains highly regular and symmetric.

\section{Conclusion \label{sec:5}}
In this work, we studied the charging performance of Heisenberg spin chain quantum batteries (QBs) under the influence of Dzyaloshinskii--Moriya interaction (DMI). We compared three distinct spin chain configurations- open boundary, closed boundary, and supercube- under both Ising and XXZ Hamiltonians. Among these, the supercube consistently exhibited superior performance. For the Ising model, DMI enhanced ergotropy in supercube spin chain. In the XXZ model, supercube achieved a stable, near-ideal sinusoidal charging--discharging cycle, enabling periodic charging with negligible residual energy per cycle at optimal interaction strengths $D=1.7$ and $J=1$. We also examined the role of the battery Hamiltonian $H_B$ in the charging process through the parameter $\lambda$. For all configurations and models, increasing $\lambda$ reduced the peak ergotropy. Interestingly, in the XXZ model, ergotropy for $D=0$ and $\lambda > 0$ remained higher than the $\lambda=0$ case for most of the evolution.  \par

Beyond these baseline comparisons, we investigated structural modifications of the supercube. Adding extra interactions between qubits disrupted the clean oscillatory behavior observed in the XXZ model, preventing full discharge to the ground state and leading to residual stored energy. To address this, we systematically extended our study to $c$-regular graphs with system sizes from $3$ to $12$ qubits. Within this framework, we analyzed both generic connectivities and highly symmetric geometries, including the tetrahedron, octahedron, and icosahedron. Remarkably, all of these structures displayed ideal sinusoidal ergotropy cycles at suitable DMI strengths. In particular, the icosahedral geometry- the largest and most symmetric case studied- stood out by fully eliminating residual energy and reaching the maximum ergotropy, thereby recovering the efficiency lost in less symmetric extensions.  \par
Although entanglement often influences QB performance, prior work has shown that entanglement generation is not strictly necessary for optimal work extraction~\cite{Hovhannisyan2013}. Consistent with this, our study did not explicitly analyze entanglement, yet demonstrated substantial enhancements in ergotropy and power without invoking it. Overall, our findings provide a roadmap for designing scalable and efficient quantum batteries. Spin chain architectures with symmetric connectivities- including the supercube, Platonic solids, and other $c$-regular graphs- achieve high-efficiency, near-ideal charging cycles without requiring entanglement, offering a promising route toward next-generation quantum energy storage devices.

\section*{Code and data availability}
The used codes and data sets are available in the \href{https://github.com/Suprabha-b/QB-HS-DMI}{GitHub} repository. The codes provided use the QuantumToolbox package in Python and Julia. \cite{QuantumToolbox-jl2025, lambert2024qutip}.

\appendix

\section{Battery performance using energy level population dynamics \label{A1}}

\begin{figure*}[h!]
    \centering

    \subfigure[Population dynamics of 8-qubit spin chains in Ising model
    \label{fig:ising pop}]
    {\includegraphics[width=0.78\linewidth]{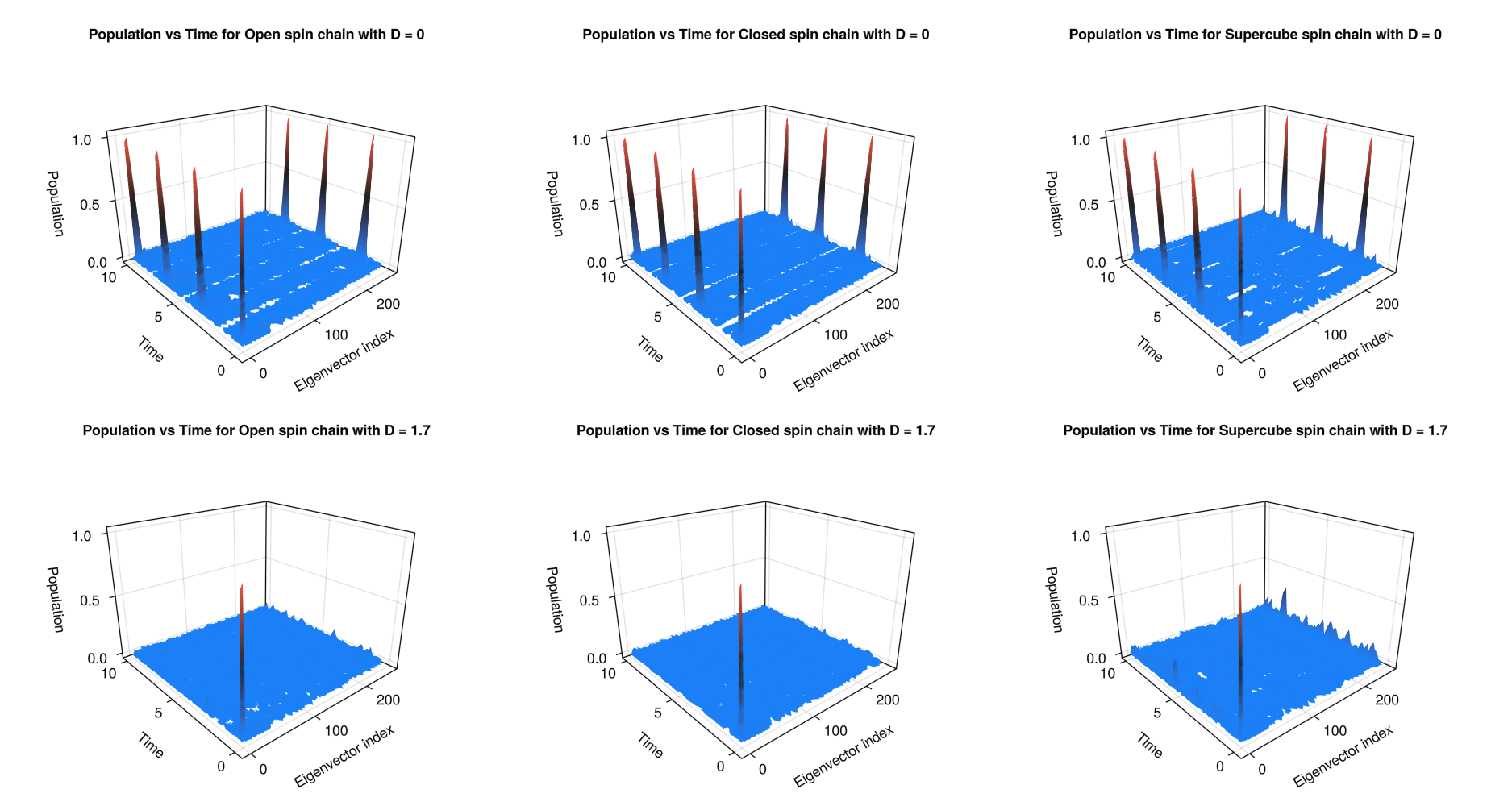}}

    \vspace{0.1cm}

    \subfigure[Population dynamics of 8-qubit spin chains in XXZ model
    \label{fig:xxz pop}]
    {\includegraphics[width=0.78\linewidth]{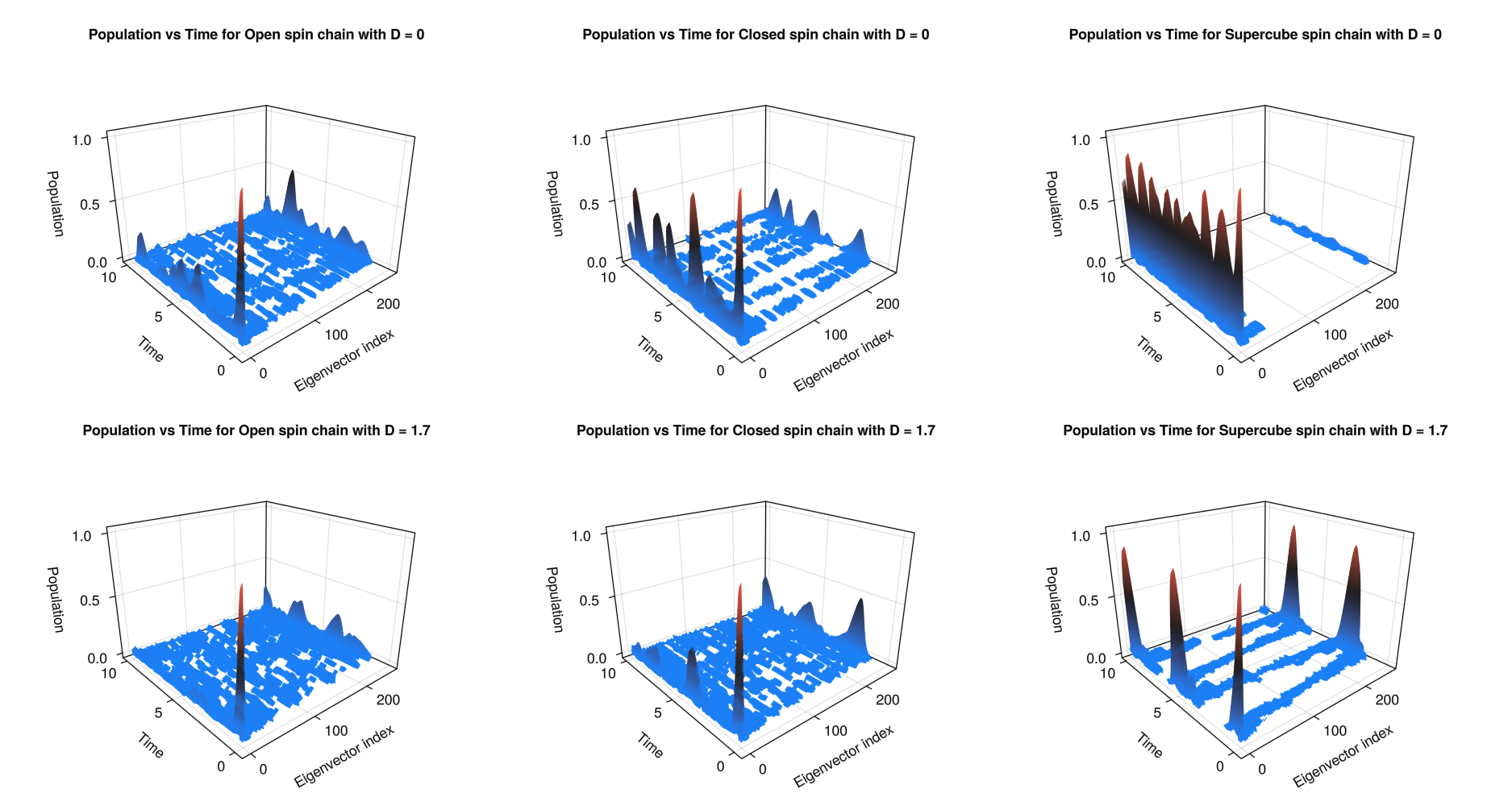}}

    \vspace{0.1cm}

    \subfigure[Population dynamics of tetrahedron QB network in XXZ model
    \label{fig:tetrahedron}]
    {\includegraphics[width=0.42\linewidth]{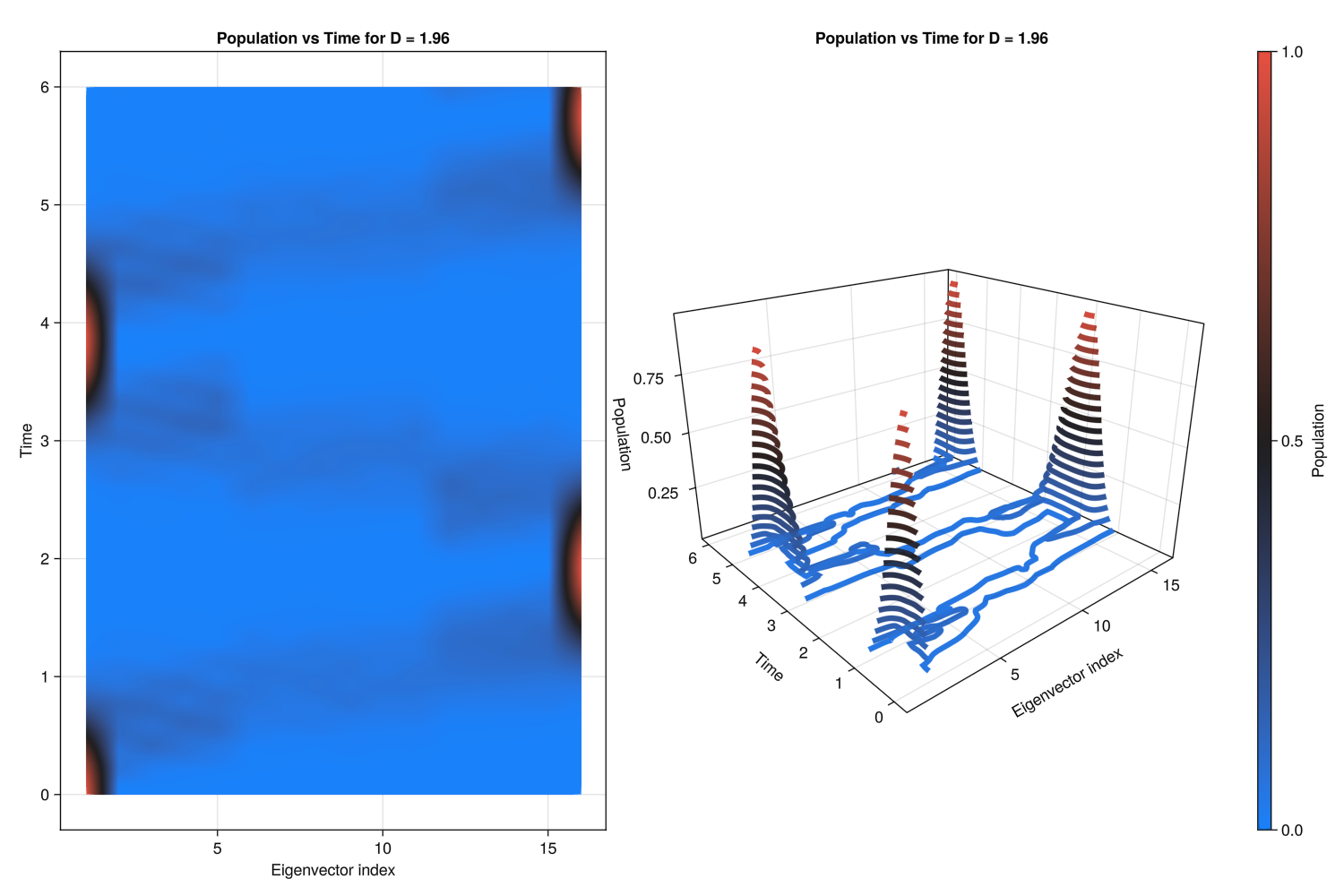}}
    \hfill
    \subfigure[Population dynamics of ground and highest excited eigenstates
    of the tetrahedron QB in XXZ model
    \label{fig:line tetrahedron}]
    {\includegraphics[width=0.42\linewidth]{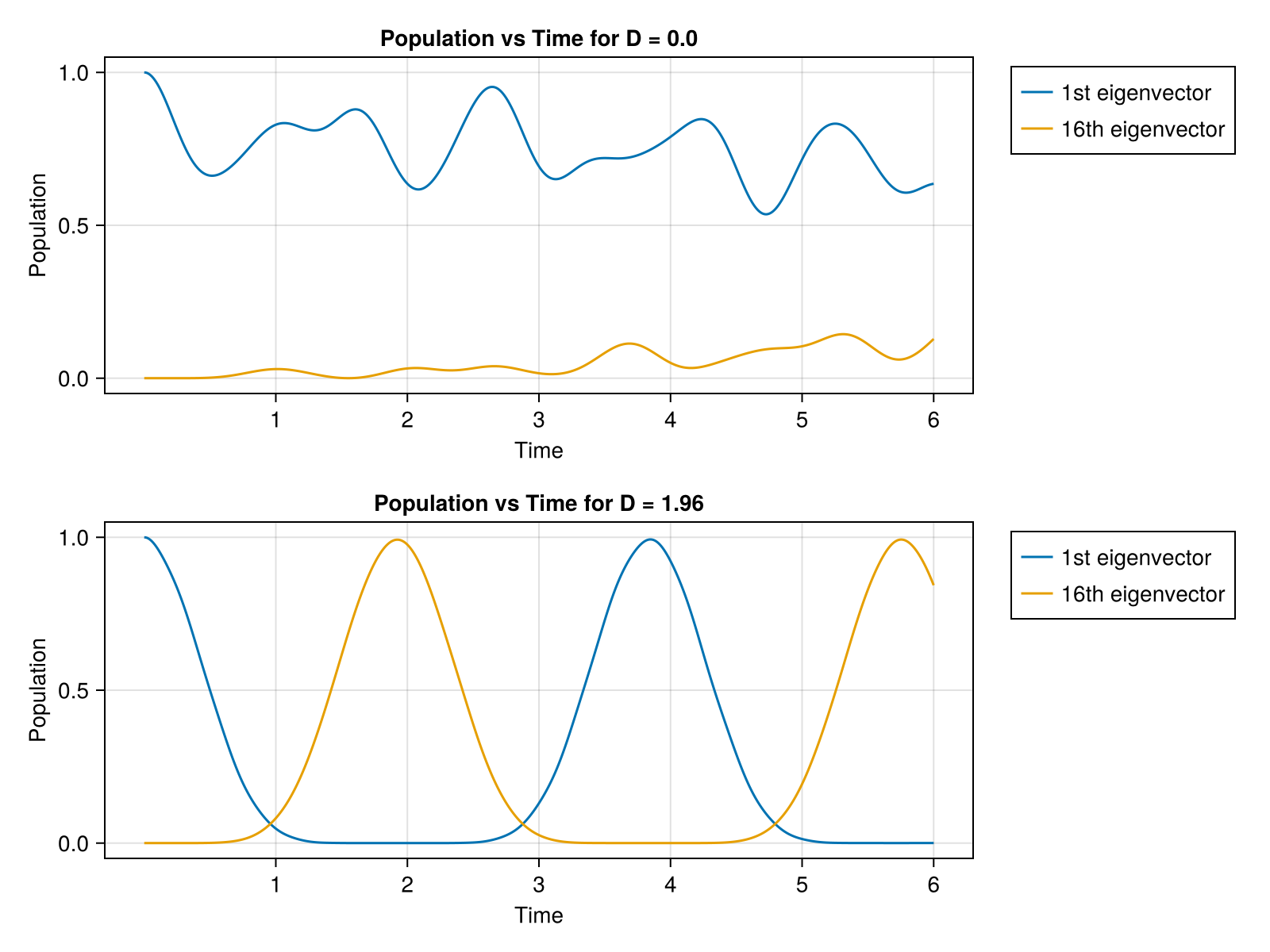}}

    \caption{
    Population dynamics of quantum battery (QB) networks.
    Top: Open boundary, closed boundary, and supercube spin chains with
    $D=0, 1.7$ for (a) Ising and (b) XXZ models.
    Bottom: (c) Tetrahedron geometry and (d) comparison between ground and
    highest excited eigenstates in the XXZ model.
    }
    \label{fig:qb_population_dynamics}
\end{figure*}


\textcolor{black}{In this section, we examine the population dynamics of the three 8-qubit QB configurations, namely, open boundary, closed boundary, and supercube spin chains, as well as the tetrahedron QB network, to provide additional insight into the geometries discussed in Section \ref{3.b.4}. Fig. \ref{fig:ising pop} and \ref{fig:xxz pop} shows the time evolution of population across eigenstates for the 8-qubit configurations, and Fig. \ref{fig:tetrahedron} shows the corresponding behavior for the tetrahedron network. In the 3D plots, the ground state is indexed as the 1st eigenvector and the highest excited state as the $(2^n)^\mathrm{th}$ eigenvector along the x-axis.}

\textcolor{black}{Population inversion in QB systems is analogous to optical pumping in lasers. Recent studies show that dissipative processes can drive quantum systems to equilibrium states featuring population inversion, where the charged state can be maintained without continuous driving once equilibrium is reached \cite{barra2019dissipative}. This behavior is especially relevant for atom–cavity QED systems, where laser-driven three-level atoms interacting with cavities in the dispersive regime achieve inversion through engineered dissipation \cite{beleno2024laser}. Analyzing eigenstate population evolution in our QB networks therefore provides deeper understanding of charging routes and energy-storage efficiency in multi-qubit systems.}

\textcolor{black}{In the Ising model without DMI, excitations from the ground state are broadly distributed: all eigenstates except the ground and highest excited states become equally populated at intermediate times, reflected in the nearly uniform blue surface in Fig. \ref{fig:ising pop}. The ground state and the highest excited state show periodic peaks, but population spreading over intermediate levels prevents smooth sinusoidal ergotropy oscillations, even though partial inversion appears. When DMI is introduced, this behavior becomes even more uniform, destroying selectivity entirely. The periodicity of peaks in the ground and highest excited states disappears, and the system no longer reaches the excited state, causing strong suppression of ergotropy oscillations as seen in Fig. \ref{fig:diff d erg ising}. This behavior is consistent across all three geometries, indicating that the Ising model is insensitive to connectivity patterns.}

\textcolor{black}{In contrast, the XXZ model without DMI never fully excites the system to its highest energy state, as shown in Fig. \ref{fig:xxz pop}. Open and closed chains show population spreading over many eigenstates. Remarkably, in the supercube configuration the population remains strongly concentrated in the ground state, indicating that its connectivity restricts spreading. Upon introducing DMI, open and closed chains continue to show broad population distribution with no pronounced peaks, consistent with the absence of oscillatory ergotropy. Strikingly, the supercube becomes the only configuration that exhibits clear periodic population inversion, with sharp, regularly spaced peaks alternating between the ground and highest excited states, while the intermediate eigenstates remain weakly populated. This selective excitation pathway reproduces the oscillation pattern in Fig. \ref{fig:erg XXZ}(c), highlighting the unique periodic energy-transfer mechanism supported by the supercube.}

\textcolor{black}{We further examine the tetrahedron QB network (a 4-qubit interaction graph) in the XXZ model to determine whether similar selective periodic inversion appears in other geometric configurations presented in Section \ref{3.b.4}. As with the supercube for $D=1.7$, the tetrahedron shows clear and regular population inversion, with sharp periodic peaks corresponding to the ground and highest excited states, as seen in Fig. \ref{fig:tetrahedron}. Weakly populated intermediate eigenstates again indicate a well-defined excitation pathway. Fig. \ref{fig:line tetrahedron} shows complete inversion via periodic peaks of the 1st and 16th eigenvectors. Similar population-inversion behavior is observed in the other graph networks listed in Table \ref{Table: geometries} and shown in Fig. \ref{fig:4q6q12q}.}


\bibliography{article}

\end{document}